\documentclass[12pt]{article}
\usepackage{graphicx}
\usepackage{bm}

\begin{document}

\title{Wigner functions, negativity volumes, and experimental generation of Pegg-Barnett phase-operator eigenstates}

\author{Hiroo Azuma\thanks{Email: hiroo.azuma@oist.jp, ORCID: 0000-0002-4374-8727}
\\
\\
{\small OIST Center for Quantum Technologies,}\\
{\small Okinawa Institute of Science and Technology,}\\
{\small 1919-1 Tancha, Onna-son, Okinawa 904-0495, Japan}\\
}

\date{\today}

\maketitle

\begin{abstract}
In this paper, we study the non-Gaussianity of the eigenstates of the
\\
Pegg-Barnett phase observable. 
By computing the Wigner functions of the eigenstates, we confirm that they take negative values in specific regions of the phase space. 
The Pegg-Barnett phase-operator eigenstates are defined in a
\\
finite-dimensional Hilbert space.
Thus, we examine how their negativity volumes depend on the dimension of the Hilbert space. 
Moreover, we present a quantum-optical circuit that generates these eigenstates and identify single-photon detection as the origin of their non-Gaussianity.
To investigate a more realistic experimental implementation,
we introduce imperfect single-photon detectors with non-unit efficiency into the circuit. We then examine how the detection probability, the output-ideal fidelity, and the negativity volume of the approximate eigenstate output from the circuit change with the detector efficiency.
Finally, as a practical application, we consider a phase-estimation experiment of an arbitrary unknown state by injecting both the unknown state and a known Pegg-Barnett eigenstate into a 50-50 beam splitter and individually counting the numbers of photons emitted from its two output ports.
\end{abstract}

\section{\label{section-introduction}Introduction}

Constructing a phase operator has long been a difficult problem in quantum mechanics.
Although Dirac attempted to introduce a phase operator for photons by analogy with the Poisson brackets of classical equations of motion \cite{Dirac1927}, the resulting operator was not Hermitian.
Susskind and Glogower later showed the non-existence of a Hermitian phase operator for photons \cite{Susskind1964}.
As an alternative approach, they defined non-commutative sine and cosine operators, from which they constructed phase-number uncertainty relations.
Following this line of reasoning, a method for constructing an approximate Hermitian phase operator was explored.
A resolution to this problem was provided by the Pegg-Barnett phase operator, which is defined in a finite-dimensional Hilbert space \cite{Pegg1988,Barnett1989,Pegg1989}.

As applications of the Pegg-Barnett phase-operator formalism, the following issues have been investigated:
the fluctuation of the Pegg-Barnett phase operator in a coherent state \cite{Lynch1990},
the fluctuation of the phase operator in a squeezed state \cite{Vaccaro1989},
distributions as functions of the phase of a cavity field that evolves under the Jaynes-Cummings interaction with an atom and the thermal effects on these distributions \cite{Eiselt1991},
and the relationship between the phase operator and the discrete Fourier transform of the number operator \cite{Perez-Leija2016}.

Strictly speaking, we cannot prepare a quantum state with a well-defined phase because no rigorous Hermitian phase operator exists.
As mentioned above, this fact was proved by Susskind and Glogower \cite{Susskind1964}.
Some might argue that a coherent state possesses a well-defined phase. 
However, the phase of a coherent state is merely a classical parameter of electromagnetism and should not be interpreted as a quantum-mechanical phase observable. 
As an approximate substitute for an ideal phase-defined state, we consider the eigenstates of the Pegg-Barnett phase operator.
These eigenstates form an orthonormal basis with well-defined phase in a finite-dimensional Hilbert space.
However, if we let the dimension of the Hilbert space be infinity, those states cannot be ideal phase-defined states \cite{Pegg1988,Pegg1989}.
Each Pegg-Barnett phase-operator eigenstate is a superposition of photon-number states with equally weighted amplitudes. 
Because photon-number states are non-Gaussian, we can expect that the eigenstates inherently exhibit non-Gaussianity.

In Ref.~\cite{Sarkar2021}, the Hermitian number and phase operators are constructed in finite-dimensional space by following the Pegg-Barnett formalism for analyzing the quantum hypergraph states.
Constructing a displacement operator from the cyclic Pegg-Barnett phase operator and its adjoint instead of annihilation and creation operators of photons, the authors of Ref.~\cite{Mouhaoui2024} defined the Pegg-Barnett coherent states and examined their statistical properties.
These Pegg-Barnett coherent states can be regarded as an alternative version of approximate quantum states with a well-defined phase.
However, we do not consider these states in the present paper.

Non-Gaussianity of a pure (i.e., non-mixed) quantum state means that its Wigner function takes negative values in some region of phase space \cite{Walls2008, Barnett1997, Wigner1932, Hillery1984}.
For example, coherent states and squeezed states are Gaussian while, as mentioned above, photon-number states are non-Gaussian.
In this sense, the non-Gaussianity of photonic states can be regarded as a property that stands in contrast to classical behavior.
However, non-Gaussianity is a concept that is different from quantum entanglement.
This can be understood from the fact that when squeezed light is split into two modes by a 50-50 beam splitter, entanglement is generated even though squeezed light itself is Gaussian.

There is no consensus on how to quantify non-Gaussianity.
One of the simplest measures is the negativity volume,
defined as the integral of the negative part of the Wigner function over phase space
\cite{Kenfack2004, Arkhipov2018}.
For continuous-variable states defined on the infinite-dimensional Hilbert space, it is known that the Wigner function of a pure state is non-negative if and only if the state is Gaussian \cite{Hudson1974}.
Thus, any pure non-Gaussian state necessarily has regions in phase space where the value of its Wigner function is negative.
Therefore, the negativity volume of a non-Gaussian state is always positive while that of a Gaussian state must be equal to zero.

Because the eigenstates of the Pegg-Barnett phase operator reside in a
\\
finite-dimensional Hilbert space,
we cannot use the negativity volume as a rigorous measure of the non-Gaussianity for the eigenstates.
To overcome this defect, the following studies have been made.
In Ref.~\cite{Wootters1987}, Wootters investigated a formulation of the Wigner functions for states defined on the finite-dimensional Hilbert space.
Using Wootters' discrete Wigner function, Vaccaro and Pegg derived the number-phase Wigner functions of the eigenstates of the Pegg-Barnett phase operator \cite{Vaccaro1990}.
However, in this paper, we adopt the negativity volume for the infinite-dimensional space to quantify the non-Gaussianity of the eigenstates of the Pegg-Barnett phase operator instead of Wootters' formalism.
This approach has the following advantage.
If we take the limit where the dimension of the Hilbert space supporting the eigenstates tends to infinity,
the negativity volume becomes well-defined and acquires a physical meaning.
Thus, examining the behavior of the Wigner functions and the negativity volume in the limit of a large Hilbert-space dimension is important.

Non-Gaussian states are generally recognized as difficult to generate experimentally.
For example, to produce a coherent-state Schr{\"{o}}dinger-cat state, which exhibits non-Gaussianity, we typically inject squeezed light into a beam splitter and subtract a single photon with regarding the detection of this photon as a heralding signal
\cite{Dakna1997,Wenger2004,Ourjoumtsev2006,Neergaard-Nielsen2006,Wakui2007}.
This method requires a high-efficiency single-photon detector, which makes the experiment challenging.
Consequently, the eigenstates of the Pegg-Barnett phase operator, which also exhibit non-Gaussianity, are expected to be difficult to generate experimentally.
This expectation is consistent with the well-known fact that an ideal Hermitian phase operator in quantum mechanics does not exist.

Experimental generation of photonic non-Gaussian states draws a lot of attention in the field of quantum optics.
Reference~\cite{Magro2023} demonstrated generation of deterministic freely propagating photonic qubits whose Wigner function showed negativity.
The application of non-Gaussian states to quantum information processing has been actively investigated at the theoretical level as a precursor to experimental implementation.
Reference~\cite{Dahan2023} showed how to create optical cat and Gottesman-Kitaev-Preskill (GKP) states using phase matching of free electrons with optical waveguides and photonic crystals in an ultrafast transmission electron microscope.
Practical quantum computation using approximate GKP states was investigated in \cite{Tzitrin2020}.

In this paper, we compute the Wigner functions
and the corresponding negativity volumes
of the eigenstates of the Pegg-Barnett phase operator.
In particular, we examine how these quantities scale with the dimension of the Hilbert space in which the eigenstates are defined.
Our numerical results suggest that the negativity volume diverges as the dimension of the Hilbert space increases.

Moreover, we examine an experimental implementation for generating the
\\
Pegg-Barnett phase-operator eigenstates.
From this analysis, we identify single-photon detection as the origin of their non-Gaussianity.
To analyze practical experimental effects, we introduce imperfect single-photon detectors with efficiencies less than unity into a quantum-optical circuit that outputs the eigenstate.
We evaluate the probability that all single-photon detectors click, the fidelity between the experimentally generated states and the ideal eigenstates, and the negativity volumes of the states emitted from the circuit.
We investigate the dependence of these quantities on the efficiency of the detectors.

Finally, we consider phase-estimation experiments as applications of the Pegg-Barnett phase-operator eigenstates.
Phase estimation is an important problem that has been studied extensively in the literature.
For example, in Ref.~\cite{Gorecki2022}, multiple-phase quantum interferometry was investigated.
In this paper, we consider the following setting.
By injecting an arbitrary unknown state and a known eigenstate as a reference beam into a 50-50 beam splitter, and by individually counting the photons leaving its two output ports, we can analyze the resulting interference pattern and statistically estimate the phase of the unknown state.
Our proposal of this method is very simple and it has a problem that the number of trials for obtaining the statistical interference pattern becomes larger as the dimension of the Hilbert space increases.
However, because we must employ advanced mathematical techniques to solve this problem, we do not pursue it further in this paper.

This article is organized as follows.
In Sec.~\ref{section-Pegg-Barnett}, first, we give a brief review of the Pegg-Barnett phase operator and its eigenstates.
Next, we numerically compute the Wigner functions and negativity volumes of the Pegg-Barnett phase-operator eigenstates.
In Sec.~\ref{section-experimental-implementation}, we study an experimental implementation for generating the eigenstates using imperfect single-photon detectors.
In Sec.~\ref{section-application-phase-measurement}, we investigate an application of these eigenstates to a phase-measurement experiment of a given unknown state.
In Sec.~\ref{section-discussion}, we present a discussion of the results.
In Sec.~\ref{section-conclusion}, we provide conclusions.
In \ref{Appendix-A}, we explain derivations of the parameters that are used in the quantum-optical circuit shown in Sec.~\ref{section-experimental-implementation}.

\section{\label{section-Pegg-Barnett}The Wigner functions and the negativity volumes of eigenstates of the Pegg-Barnett phase operator}

In this section, we numerically compute the Wigner functions and the negativity volumes of eigenstates of the Pegg-Barnett phase operator.
First of all, we give a brief review of the Pegg-Barnett phase operator and its eigenstates.

The Pegg-Barnett phase operator is defined in an $(s+1)$-dimensional Hilbert space,
where $s (\ge 1)$ is an integer.
The phase state corresponding to $\phi_{0}$ is given by
\begin{equation}
|\phi_{0},s\rangle
=
(s+1)^{-1/2}\sum_{n=0}^{s} \exp(i n \phi_{0}) |n\rangle,
\label{phi-0-eigenstate}
\end{equation}
where $\{|n\rangle : n = 0, 1, \ldots, s\}$ denotes the Fock basis states.
Assuming
\begin{equation}
\phi_{m}
=
\phi_{0} + \frac{2\pi}{s+1} m
\quad
\mbox{for } m = 0, 1, \ldots, s,
\end{equation}
the set $\{|\phi_{m},s\rangle : m = 0, 1, \ldots, s\}$ forms an orthonormal basis of the space.
We can then construct the Pegg-Barnett phase operator as
\begin{equation}
\hat{\phi}
=
\sum_{m=0}^{s} \phi_{m} |\phi_{m},s\rangle\langle \phi_{m},s|.
\label{definition-phase-operator-0}
\end{equation}
The phase operator $\hat{\phi}$ given by Eq.~(\ref{definition-phase-operator-0}) is defined on the finite-dimensional Hilbert space
although an ideal one must live in the infinite-dimensional space.
This approximation is a characteristic of the Pegg-Barnett phase operator.
In the limit $s\to\infty$, $\hat{\phi}$ defined in Eq.~(\ref{definition-phase-operator-0})
appears to approach the ideal phase operator.
However, according to Eq.~(\ref{phi-0-eigenstate}), $|\phi_{m},s\rangle\langle\phi_{m},s|$
is proportional to $(s+1)^{-1}$, so that it converges to zero for $s\to\infty$.
Therefore, taking the limit as $s\to\infty$ is not physically meaningful and one must work with a large but finite value of $s$.

Next, we derive closed-form expressions of the Wigner functions of the eigenstates.
The Wigner function of an arbitrary density operator $\hat{\rho}$ is defined as
\begin{equation}
W(q,p)
=
\frac{1}{2\pi\hbar}
\int_{-\infty}^{\infty}
dx\,
\langle q+\frac{x}{2}|
\hat{\rho}
|q-\frac{x}{2}\rangle
e^{ipx/\hbar}.
\end{equation}
From now on, we set $\hbar = 1/2$.
By substituting $\hat{\rho} = |\phi_{m},s\rangle\langle \phi_{m},s|$ into this definition, we obtain the expression
\begin{equation}
W(q,p; s, m)
=
\frac{1}{(s+1)\pi}
\int_{-\infty}^{\infty}
dx
\sum_{k=0}^{s}\sum_{l=0}^{s}
\cos[(k-l)\phi_{m} + 2px]\,
\langle q+\frac{x}{2}|k\rangle
\langle l|q-\frac{x}{2}\rangle.
\label{Wigner-formula-phi-m}
\end{equation}
The position-representation wave functions of the Fock states are given by
\begin{eqnarray}
\langle x|0\rangle
&=&
\left(\frac{2}{\pi}\right)^{1/4}
e^{-x^{2}}, \nonumber \\
\langle x|n\rangle
&=&
\frac{1}{\sqrt{n}}
\left(
x - \frac{1}{2}\frac{d}{dx}
\right)
\langle x|n-1\rangle,
\label{position-representation-wave-functions}
\end{eqnarray}
where we use the relations for the photon creation operator,
$\hat{a}^{\dagger}|n-1\rangle = \sqrt{n}\,|n\rangle$ for $n \ge 1$
and
$\hat{a}^{\dagger} = \hat{x} - i\hat{p} = x - (1/2)(d/dx)$.

In accordance with Eqs.~(\ref{Wigner-formula-phi-m}) and (\ref{position-representation-wave-functions}), the recursive calculations yield the following results:
\begin{eqnarray}
W(q,p;1,0)
&=&
\frac{4}{\pi}
\exp[-2(q^{2}+p^{2})](q+q^{2}+p^{2}), \nonumber \\
W(q,p;2,0)
&=&
\frac{2}{3\pi}
\exp[-2(q^{2}+p^{2})]
\{1+8p^{4}+4p^{2}(-1+2q)(1+\sqrt{2}+2q) \nonumber \\
&&
+4q(1+q)[1-\sqrt{2}+2q(-1+\sqrt{2}+q)]\}, \nonumber \\
W(q,p;3,0)
&=&
\frac{2}{3\pi}
\exp[-2(q^{2}+p^{2})]
\{8p^{6}-4p^{4}[3+\sqrt{6}-2q(\sqrt{3}+3q)] \nonumber \\
&&
+p^{2}[3(2-\sqrt{2}+\sqrt{6})
+4q(-3\{\sqrt{3}+\sqrt{2-\sqrt{3}}\}
-6q+4\sqrt{3}q^{2}+6q^{3})] \nonumber \\
&&
+q[3-3\sqrt{2}+3\sqrt{3}+q(6+3\sqrt{2}-3\sqrt{6}+4q\{\sqrt{3}[-3+\sqrt{2+\sqrt{3}}] \nonumber \\
&&
+q[-3+\sqrt{6}+2q(\sqrt{3}+q)]\})]\}.
\end{eqnarray}
As these examples illustrate, the closed-form expression of $W(q,p;s,m)$ becomes significantly more complex as $s$ increases.

\begin{figure}
\begin{center}
\includegraphics[width=0.6\linewidth]{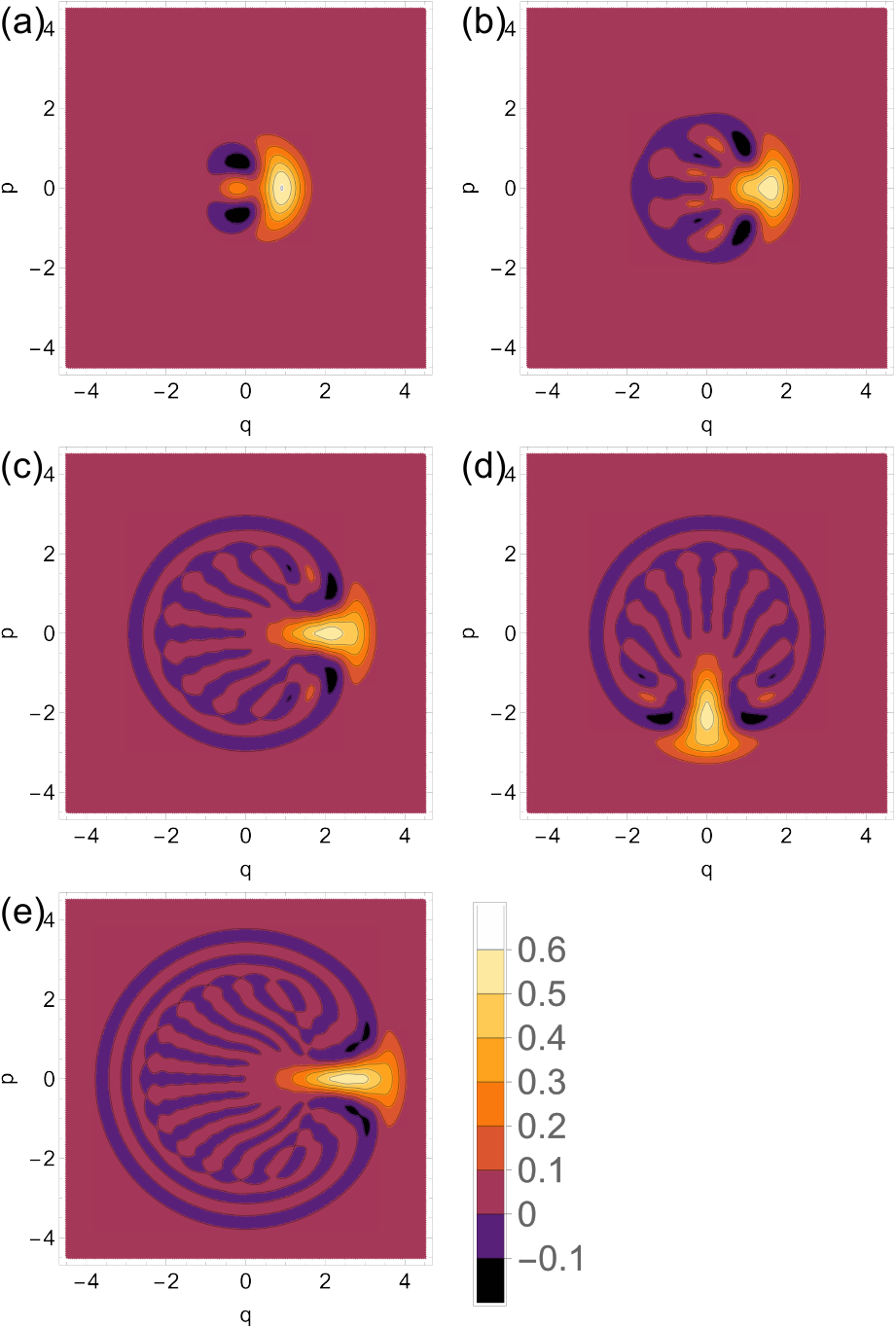}
\end{center}
\caption{Two-dimensional plots of the Wigner functions $W(q,p;s,m)$ for the states
$|\phi_{m},s\rangle$ in the $q$-$p$ plane, with $\phi_{0}=0$.
(a) $s=2$, $m=0$, (b) $s=5$, $m=0$, (c) $s=11$, $m=0$,
(d) $s=11$, $m=3$, $\phi_{3}=\pi/2$,
(e) $s=17$, $m=0$.
As $s$ increases, the interference-like ripple structures become gradually finer.
Comparing (c) and (d), we note that the pattern rotates clockwise as the phase increases
from $\phi_{0}=0$ to $\phi_{3}=\pi/2$.}
\label{figure01}
\end{figure}

\begin{figure}
\begin{center}
\includegraphics[width=0.5\linewidth]{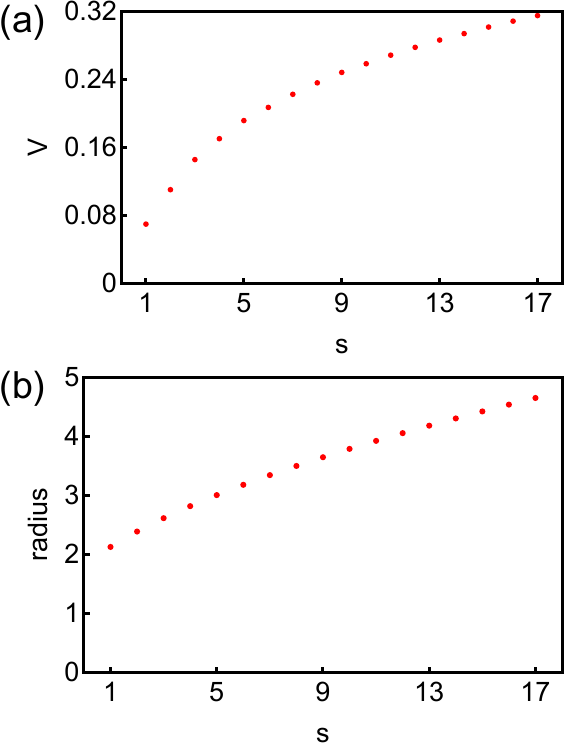}
\end{center}
\caption{(a) A plot of ${\cal V}_{s,0}$ as a function of $s$.
The quantity ${\cal V}_{s,0}$ increases monotonically with $s$.
(b) A plot of the radius of $W(q,p;s,0)$ as a function of $s$.
The radius increases monotonically with $s$.}
\label{figure02}
\end{figure}

In Figs.~\ref{figure01}(a),(b),(c),(e), we plot $W(q,p;s,0)$ as a function of $q$ and $p$ for
$s=2, 5, 11$, and $17$, assuming $\phi_{0}=0$.
In Fig.~\ref{figure01}(d), we plot $W(q,p;11,3)$, corresponding to $\phi_{3}=\pi/2$.
Across these plots, we observe a prominent ``mountain range'' and $(s-1)$ smaller ones.
These structures form a roughly circular region whose radius increases with $s$.
In Figs.~\ref{figure01}(a),(b),(c),(e), we also observe regions where $W(q,p;s,0)<0$,
indicating the non-Gaussianity of $|\phi_{0},s\rangle$.
Comparing Fig.~\ref{figure01}(c) with Fig.~\ref{figure01}(d), we see that the pattern
rotates clockwise as the phase increases.

To quantify the non-Gaussianity numerically, we use the negativity volume as a metric.
The negativity volume is defined as
\begin{equation}
{\cal V}_{s,m}
=
\frac{1}{2}
\left(
\int dq \int dp\, |W(q,p; s,m)| - 1
\right).
\end{equation}
In Fig.~\ref{figure02}(a), we plot ${\cal V}_{s,0}$ as a function of $s$.
The negativity volume tells us how much of the Wigner function takes negative values, by integrating its negative regions.
This shows how far a state is from behaving like a classical probability distribution in phase space.
Since the Wigner functions of the Pegg-Barnett phase-operator eigenstates have distinctive nonclassical patterns, the negativity volume is a natural and informative way to characterize them.

In Fig.~\ref{figure02}(a), we examine how the negativity volume changes as the parameter $s$ increases, i.e., as the dimension of the Hilbert space is extended
and we note that ${\cal V}_{s,0}$ increases monotonically with $s$.
Increasing $s$ enlarges the orthonormal basis $\{|\phi_{m},s\rangle\}$,
and using a large set of basis vectors allows phase measurements with higher resolutions.
For this reason, it is practically important to investigate the behavior of the eigenstates in the large-$s$ limit.

We have explained in Fig.~\ref{figure01} that the radius of the roughly circular region of
$W(q,p;s,0)$ increases with $s$.
We now verify this behavior quantitatively.
We define the radius of $W(q,p;s,0)$ as follows.
Plotting the one-dimensional function $W(q,0;s,0)$ versus $q$,
we identify the point $q_{0}$ that satisfies $W(q_{0},0;s,0)=0.001$ and
$|W(q,0;s,0)|<0.001$ for $q>q_{0}$.
Then, we regard $q_{0}$ as the radius.
Next, we plot the radius as a function of $s$ in Fig.~\ref{figure02}(b).
Looking at Fig.~\ref{figure02}(b), we note that the radius increases monotonically with $s$.

If the dimension of the Hilbert space is extended to infinity, the phase space becomes continuous and the Wigner function recovers its standard definition as a quasiprobability distribution.
In this limit, the negativity volume acquires a clear physical meaning, because the integrated weight of the negative regions directly quantifies the nonclassicality of the state in the usual continuous phase space.
In contrast, for finite $s$, the phase space is discretized and the interpretation of the negativity volume is less straightforward.
Therefore, examining the behavior of the negativity volume in the large-$s$ limit is essential for understanding how the Pegg-Barnett eigenstates connect to the ideal phase operator framework.

As shown in Figs.~\ref{figure02}(a) and (b), both the negativity volume and the radius appear to diverge as $s$ increases.
This behavior indicates that, in the limit $s\to\infty$, the Pegg-Barnett phase-operator eigenstates would become exceedingly difficult to generate experimentally.
This observation is consistent with the fact that no exact Hermitian phase operator exists.

\section{\label{section-experimental-implementation}Experimental implementation of Pegg-Barnett
\\
phase-operator eigenstates}

As discussed in Sec.~\ref{section-Pegg-Barnett}, the eigenstates of the Pegg-Barnett phase operator exhibit non-Gaussianity.
In this section, we investigate an experimental method for generating these eigenstates and identify the origin of their non-Gaussianity.
From this analysis, we find that the non-Gaussianity emerges from single-photon detection.

\begin{figure}
\begin{center}
\includegraphics[width=0.5\linewidth]{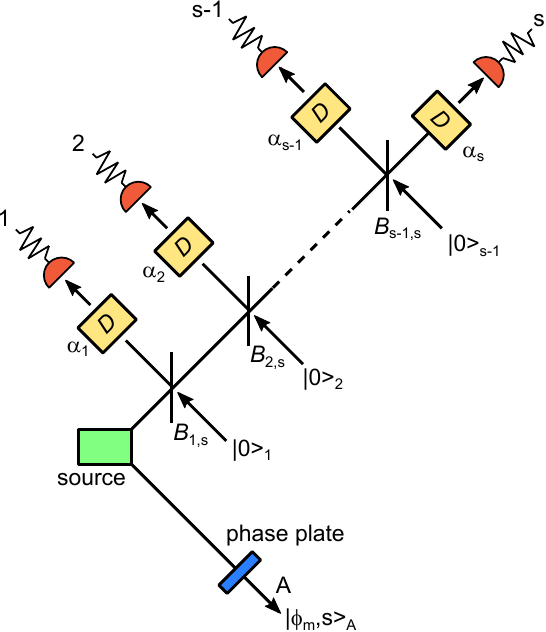}
\end{center}
\caption{A schematic illustration of the implementation for the approximate generation of
$|\phi_{m},s\rangle$.
The procedure is as follows.
A source emits a two-mode squeezed vacuum state in modes $s$ and A.
Vacuum states $|0\rangle_{1}|0\rangle_{2}\cdots|0\rangle_{s-1}$ are prepared in modes $1,2,\ldots ,s-1$, and the beam splitters
$\hat{B}_{1,s}$, $\hat{B}_{2,s}$, ..., $\hat{B}_{s-1,s}$ are applied to modes $1, 2, ..., s$.
Then, the displacement operators $\hat{D}_{j}(\alpha_{j})$ ($j=1, 2, ..., s$) are applied to these modes.
Single-photon detection is individually performed on modes $1, 2, ..., s$.
Finally, a phase plate is placed on mode A to obtain $|\phi_{m},s\rangle$ approximately, with $\phi_{0}=0$.}
\label{figure03}
\end{figure}

In Fig.~\ref{figure03}, we show a quantum-optical circuit that generates the
eigenstate of the Pegg-Barnett phase operator, $|\phi_{m},s\rangle$.
The circuit works as follows.
First, a source emits a two-mode squeezed vacuum state in modes $s$ and A, and we
append vacuum states $|0\rangle_{1}|0\rangle_{2}\cdots|0\rangle_{s-1}$ in modes $1, 2, \ldots, s-1$ to prepare the initial state,
\begin{equation}
\sqrt{1-q^{2}}
\sum_{n=0}^{\infty} q^{n}
|0\rangle_{1}
|0\rangle_{2}
\cdots
|0\rangle_{s-1}
|n\rangle_{s}
|n\rangle_{\mathrm{A}},
\end{equation}
where $q = \tanh r$ and $r$ is the squeezing parameter.
Second, we sequentially apply the beam splitters
$\hat{B}_{1,s}, \hat{B}_{2,s}, \ldots, \hat{B}_{s-1,s}$
to modes
$1, 2, \ldots, s$, where each beam splitter is defined by
\begin{equation}
\hat{B}_{k,s}:
\left(
\begin{array}{c}
\hat{a}'_{k} \\
\hat{a}'_{s} \\
\end{array}
\right)
=
\left(
\begin{array}{cc}
\sqrt{(s-k)/(s-k+1)} & -1/\sqrt{s-k+1} \\
1/\sqrt{s-k+1} & \sqrt{(s-k)/(s-k+1)} \\
\end{array}
\right)
\left(
\begin{array}{c}
\hat{a}_{k} \\
\hat{a}_{s} \\
\end{array}
\right),
\end{equation}
for $1 \le k \le s-1$.
Third, for $|\alpha_{j}| \ll 1$, we apply the displacement operators
\begin{equation}
\hat{D}_{j}(\alpha_{j})
\simeq
1
+\alpha_{j}\hat{a}_{j}^{\dagger}
- \alpha_{j}^{*}\hat{a}_{j}
\quad
\mbox{for } j = 1, 2, \ldots, s,
\end{equation}
where we choose $\{\alpha_{j}\}$ and $q$ to be small quantities of the same order.
To adjust the parameters $\{\alpha_{j}\}$, we express them as functions of $q$.
This procedure is detailed in \ref{Appendix-A}.
We explain how to adjust $q$ in Figs.~\ref{figure04}, \ref{figure05}, and \ref{figure06}.

From now on, as a concrete example, we set $s=4$.
Consequently, the state of mode A becomes an equal-amplitude superposition,
$|\varphi_{4}\rangle+O(q)$,
where
\begin{equation}
|\varphi_{4}\rangle
=
|\phi_{0},4\rangle
=
\frac{1}{\sqrt{5}}
(|0\rangle + |1\rangle + |2\rangle + |3\rangle + |4\rangle).
\label{definition-ket-varphi4}
\end{equation}
Then, we place a phase plate whose Hamiltonian is given by
$\hat{H}=\hbar\omega\hat{a}^{\dagger}\hat{a}$ on mode A,
where $\omega$ denotes the angular frequency of photons in mode A.
After this operation, the superposition transforms into
\begin{equation}
(s+1)^{-1/2}
\sum_{n=0}^{s} \exp(-in\omega t)\, |n\rangle + O(q),
\label{final-approximate-state}
\end{equation}
where $s = 4$.
By adjusting the interaction time such that $\phi_{m} = -\omega t$,
we obtain an approximate realization of the state $|\phi_{m},s\rangle$.
In the above method, we must solve an $s$-th-degree equation.

Here, we examine the optical circuit shown in Fig.~\ref{figure03} carefully.
First, the source emits the two-mode squeezed vacuum state into modes $s$ and A.
It is well known that we can generate the two-mode squeezed vacuum state by injecting two squeezed states separately onto a 50-50 beam splitter.
Because the squeezed states are Gaussian and the beam splitter is a linear optical device, we can regard this first step as Gaussian.
Second, we prepare the vacuum states $|0\rangle_{1}|0\rangle_{2}...|0\rangle_{s-1}$ and apply the beam splitters to modes $1, 2, ..., s$.
This second step includes only Gaussian elements.
Third, we apply the displacement operators to modes $1, 2, ..., s-1$.
Because the displacement operator is generated by a Hamiltonian that is linear in the annihilation and creation operators,
it is classified as a Gaussian operation.
Fourth, we perform single photon detections in modes $1, 2, ..., s$.
Because photon-number states are well-known non-Gaussian states, the fourth step involves a non-Gaussian operation.
Finally, we apply the phase plate to mode A.
Because the phase plate is a linear optical device, this operation is Gaussian.
Summarizing the above, in the quantum-optical circuit shown in Fig.~\ref{figure03}, 
the only non-Gaussian element is the single-photon detection in modes
$1, 2, \ldots, s$.
Thus, the non-Gaussianity of the state $|\phi_{m},s\rangle$ emerges solely from
the single-photon detection.
Here, we emphasize the following point.
The quantum-optical circuit shown in Fig.~\ref{figure03} was originally developed
to generate a superposition of photon-number states with tunable coefficients
\cite{Lvovsky2002,Bimbard2010,Marek2011,Yukawa2013,Azuma2026}.

To consider a more realistic experimental implementation, we assume imperfect
photon detectors described by the positive-operator-valued measure (POVM)
$\{\hat{E},\, \hat{\bm{I}} - \hat{E}\}$, where
\begin{equation}
\hat{E}
=
\eta
\sum_{k=1}^{\infty} (1-\eta)^{k-1} |k\rangle\langle k|,
\end{equation}
for a ``click'', $\hat{\bm{I}} - \hat{E}$ for ``no click'',
and $\eta$ denotes the detection efficiency
\cite{Kok2007,Resch2001,Akhlaghi2011}.
Here, we focus on the following two quantities to evaluate the performance of the circuit shown in Fig.~\ref{figure03}:
(i) the probability that all four imperfect detectors click  simultaneously,
\begin{equation}
P = \langle\Psi|
(\hat{E}\otimes\hat{E}\otimes\hat{E}\otimes\hat{E}\otimes\hat{\bm{I}})
|\Psi\rangle,
\end{equation}
where $|\Psi\rangle$ is given by Eq.~(\ref{state-circuit-output-0});
and (ii) the fidelity between the post-selected output state and the ideal
equal-amplitude superposition,
\begin{equation}
F = \frac{1}{P}
\langle\Psi|
(\hat{E}\otimes\hat{E}\otimes\hat{E}\otimes\hat{E}\otimes
|\varphi_{4}\rangle\langle\varphi_{4}|)
|\Psi\rangle,
\end{equation}
where $|\varphi_{4}\rangle$ is given by Eq.~(\ref{definition-ket-varphi4}).

\begin{figure}
\begin{center}
\includegraphics[width=0.5\linewidth]{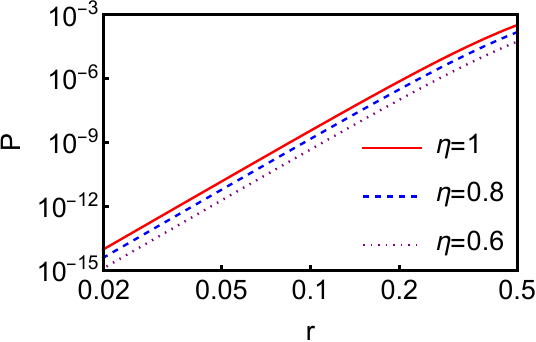}
\end{center}
\caption{Plots of $P$, the probability that all four imperfect detectors click simultaneously for $s=4$, as a function of $r = \mathrm{arctanh}\, q$.
Both the horizontal and vertical axes are displayed by the logarithmic scale.
The solid red, dashed blue, and dotted purple curves represent
$\eta = 1$, $0.8$, and $0.6$, respectively.
We observe from these curves that $P$ depends only weakly on $\eta$.
All these curves can be well fitted by
$\log_{10}P=c+8\log_{10}r$, where $c$ is a function of $\eta$.}
\label{figure04}
\end{figure}

\begin{figure}
\begin{center}
\includegraphics[width=0.5\linewidth]{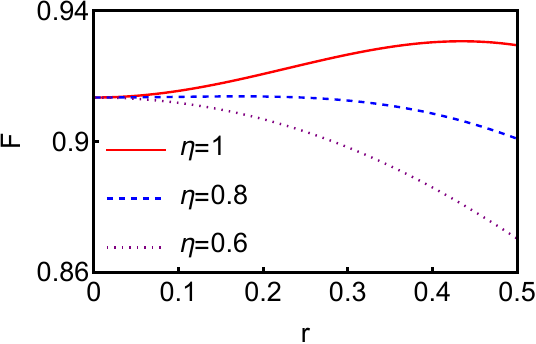}
\end{center}
\caption{Plots of $F$, the fidelity between the post-selected output state and the
equal-amplitude superposition, as a function of $r = \mathrm{arctanh}\, q$ for $s=4$.
The solid red, dashed blue, and dotted purple curves correspond to
$\eta = 1$, $0.8$, and $0.6$, respectively.
From these curves, we note that $F$ depends strongly on $\eta$.}
\label{figure05}
\end{figure}

In Fig.~\ref{figure04}, we plot $P$ as a function of
$r = \mathrm{arctanh}\, q$ for $\eta = 1$, $0.8$, and $0.6$.
From Eqs.~(\ref{projected-state-mode-A-0}), (\ref{coefficients-projected-state-mode-A-0}), and (\ref{relations-coefficients-projected-state-mode-A-0}),
we find that $P\sim O(q^{8})\sim O(r^{8})$.
Figure~\ref{figure04} confirms this scaling behavior.
By generalizing this result, we conclude that $P\sim O(r^{2s})$ for arbitrary $s(\geq 1)$.
In other words, the probability decreases rapidly with increasing $s$.
This fact represents a significant drawback of the quantum-optical circuit shown in Fig.~\ref{figure03}.
We observe from Fig.~\ref{figure04} that $P$ is only weakly affected by $\eta$.
In Fig.~\ref{figure05}, we plot $F$ as a function of $r$ for the same values of $\eta$.
In contrast to $P$, we see from Fig.~\ref{figure05} that $F$ depends strongly on $\eta$.
The curves for $\eta=1$, $\eta=0.8$, and $\eta=0.6$
reach local maxima at $r=0.4346$ with $F=0.9307$, at $r=0.1716$ with $F=0.9139$, and in the limit $r\to\infty$ with $F=0.9135$, respectively.
Thus, $F\leq 0.9307$ holds for arbitrary $r$ and $\eta$.
The reason why the fidelity $F$ does not reach unity can be explained by Eq.~(\ref{final-approximate-state}).
According to Eq.~(\ref{final-approximate-state}),
the eigenvectors of the approximate Pegg-Barnett phase operator generated by the method shown in Fig.~\ref{figure03} contain an error of order $O(q)$,
that is, first-order error in $q$.
Consequently, the fidelity also acquires an error of order $q=\tanh r$.
For the numerical computations of $P$ and $F$ shown in
Figs.~\ref{figure04} and \ref{figure05},
we approximate the initial two-mode squeezed vacuum state and the displacement
operator by truncating the sums to the first six terms as
$\sqrt{1-q^{2}}\sum_{n=0}^{5} q^{n} |n\rangle_{4} |n\rangle_{\mathrm{A}}$
and
$\hat{D}(\alpha) \simeq \sum_{n=0}^{5} (1/n!)
(\alpha \hat{a}^{\dagger} - \alpha^{*} \hat{a})^{n}$,
respectively.

\begin{figure}
\begin{center}
\includegraphics[width=0.5\linewidth]{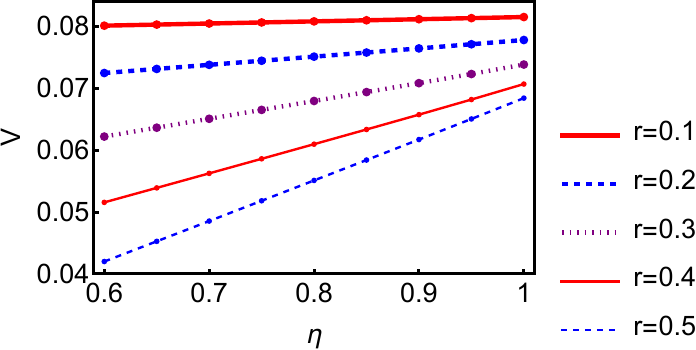}
\end{center}
\caption{Plots of ${\cal V}$, the negativity volume of the approximate state emitted
as $|\phi_{0}\rangle_{4}$ in mode A of the quantum-optical circuit shown in
Fig.~\ref{figure03} for $s=4$.
For fixed squeezing parameter $r$, we plot ${\cal V}$ as a function of $\eta$,
the efficiency of the single-photon detectors.
The thick solid red, thick dashed blue, thick dotted purple, thin solid red, and thin dashed blue curves represent
$r = 0.1$, $0.2$, $0.3$, $0.4$, and $0.5$, respectively.}
\label{figure06}
\end{figure}

In Fig.~\ref{figure06}, we plot ${\cal V}$, the negativity volume, for the approximate state
generated as $|\phi_{0},4\rangle$ in mode A by the quantum-optical circuit
shown in Fig.~\ref{figure03}.
We observe from Fig.~\ref{figure06} that ${\cal V}$ increases monotonically with $\eta$ for fixed $r$.
Moreover, ${\cal V}$ decreases monotonically with $r$ for fixed $\eta$.

Here, we discuss the experimental implementation of the optical circuit shown in Fig.~\ref{figure03}, since the scheme in Fig.~\ref{figure03} represents merely a conceptual circuit designed for numerical simulation.
In Fig.~\ref{figure04} and \ref{figure05}, we assume that the squeezing parameter $r$ lies in the range $0\leq r\leq 0.5$.
This assumption is reasonable, as experimental demonstrations of squeezed-state generation have achieved
3.09 dB ($r=0.356$) \cite{Zhao2020} and even
10.5 dB ($r=1.21$) \cite{Gao2024} of quadrature squeezing.
Implementation of the beam splitters, weak displacement operations, and the phase plate is practically feasible, since these are linear optical operations.
The most difficult elements in the circuit are single-photon detectors.
In Figs.~\ref{figure04}, \ref{figure05}, and \ref{figure06}, we assume the efficiency of the single-photon detection is larger than $0.6$.
Chang {\it et al} \cite{Chang2021} demonstrated a single-photon detector with efficiency in the range $0.94\leq\eta\leq 0.98$ at wavelengths of $1260$-$1625$ [nm] and timing trigger $15$-$26$ [ps].
Thus, the values of $\eta$ used in Figs.~\ref{figure04}, \ref{figure05}, and \ref{figure06} are reasonable.
Taking the above considerations into account, we conclude that the schematic illustration in Fig.~\ref{figure03} is practically feasible.

\section{\label{section-application-phase-measurement}
Application of the eigenstates of the phase operator to a phase-measurement experiment}

The phase estimation of an arbitrary photonic state is a classical important problem.
To estimate a phase shift between two optical paths, we can employ the Mach-Zehnder interferometer.
The question of how to surpass the standard quantum limit of phase sensitivity has been extensively studied by many researchers \cite{Polino2020}.
Reference \cite{Najafi2023} reported phase estimation of definite photon number states by using quantum circuits.
It studied how to estimate the unknown phase shift inside a Mach-Zehnder interferometer with photon loss by simulation.
In Ref.~\cite{Genoni2011}, optical phase estimation in the presence of phase diffusion is studied for phase-shifted Gaussian states and its ultimate quantum limits were evaluated.

If the state $|\phi_{m},s\rangle$ can be generated experimentally with high
accuracy, it can be used for quantum phase measurements.
Here, we consider the following schemes for estimating the unknown phase of a
given state.

\begin{figure}
\begin{center}
\includegraphics[width=0.4\linewidth]{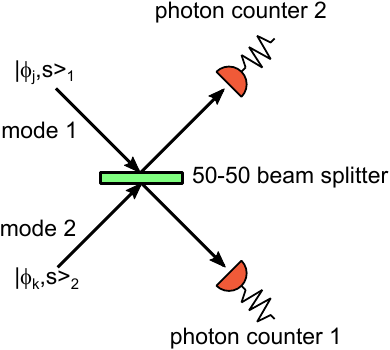}
\end{center}
\caption{A schematic illustration of the phase-estimation experiment.
The value of $\phi_{k}$ can be estimated using the reference beam
$|\phi_{j},s\rangle_{1}$.}
\label{figure07}
\end{figure}

First, we consider the problem in its simplest setting.
We prepare two states, $|\phi_{j},s\rangle$ with a known phase $\phi_{j}$ and
$|\phi_{k},s\rangle$ with an unknown phase $\phi_{k}$.
Our aim is to estimate the value of $\phi_{k}$.
We inject these two states into a 50-50 beam splitter $\hat{B}$, as shown in
Fig.~\ref{figure07}, where
\begin{equation}
\hat{B}:
\left(
\begin{array}{c}
\hat{a}'_{1} \\
\hat{a}'_{2}
\end{array}
\right)
=
\left(
\begin{array}{cc}
1/\sqrt{2} & -1/\sqrt{2} \\
1/\sqrt{2} & 1/\sqrt{2}
\end{array}
\right)
\left(
\begin{array}{c}
\hat{a}_{1} \\
\hat{a}_{2}
\end{array}
\right).
\end{equation}
Using this transformation, we obtain
\begin{equation}
\hat{B}
|\phi_{j},s\rangle_{1}
|\phi_{k},s\rangle_{2}
=
\frac{1}{s+1}
\sum_{l=0}^{s}
\sum_{m=0}^{s}
\exp[i(l\phi_{j}+m\phi_{k})]
\frac{1}{\sqrt{l!m!2^{l+m}}}
(\hat{a}_{1}^{\dagger}-\hat{a}_{2}^{\dagger})^{l}
(\hat{a}_{1}^{\dagger}+\hat{a}_{2}^{\dagger})^{m}
|0\rangle_{1}|0\rangle_{2}.
\end{equation}
The probabilities
$P(n_{1},n_{2};\phi_{j})$ of detecting $n_{1}$ and $n_{2}$ photons in modes
1 and 2, respectively, are then given by
\begin{eqnarray}
P(0,0;\phi_{j})
&=&
\frac{1}{(s+1)^{2}}, \nonumber \\
P(0,1;\phi_{j})
&=&
\frac{1}{(s+1)^{2}}
[1-\cos(\phi_{j}-\phi_{k})], \nonumber \\
P(1,0;\phi_{j})
&=&
\frac{1}{(s+1)^{2}}
[1+\cos(\phi_{j}-\phi_{k})], \nonumber \\
P(0,2;\phi_{j})
&=&
\frac{1}{(s+1)^{2}}
\left[
\frac{3}{2}-\sqrt{2}\cos(\phi_{j}-\phi_{k})
\right], \nonumber \\
P(2,0;\phi_{j})
&=&
\frac{1}{(s+1)^{2}}
\left[
\frac{3}{2}+\sqrt{2}\cos(\phi_{j}-\phi_{k})
\right], \nonumber \\
P(1,1;\phi_{j})
&=&
\frac{2}{(s+1)^{2}}
\sin^{2}(\phi_{j}-\phi_{k}), \nonumber \\
\ldots && .
\end{eqnarray}
Thus, by experimentally determining $P(n_{1},n_{2};\phi_{j})$ for
$(n_{1},n_{2})\neq (0,0)$, the unknown phase $\phi_{k}$ can be estimated.

Here, a practical difficulty arises.
Although estimating $\phi_{k}$ requires only the probability $P(0,1;\phi_{j})$
among the $[(s+1)^{2}-1]$ statistical quantities
$\{P(n_{1},n_{2};\phi_{j}) : (n_{1},n_{2}) \neq (0,0)\}$,
it is experimentally difficult to extract only this probability.
Instead, repeated trials yield the entire statistical set
$\{P(n_{1},n_{2};\phi_{j})\}$ simultaneously.
Consequently, even though only $P(0,1;\phi_{j})$ is needed, we are forced to
obtain $P(1,0;\phi_{j})$, $P(1,1;\phi_{j})$, and so on, which leads to a
substantial resource overhead when $s \gg 1$.
However, a detailed investigation of possible resolutions to this issue lies beyond the
scope of the present work, and we therefore do not pursue it further here.

Next, we design an experiment that allows us to determine the coefficients of a
superposition of $\{|\phi_{m},s\rangle\}$, namely the coefficients
$\{c_{k}\}$ in $|\psi\rangle = \sum_{k=0}^{s} c_{k} |\phi_{k},s\rangle$.
We prepare a reference state with a known phase in mode~1 as
$|\phi_{j},s\rangle_{1}$, and an unknown superposition in mode~2 as
$|\psi\rangle_{2}$.
These two states are injected into a 50-50 beam splitter, after which we count
the number of emitted photons in each output port of modes~1 and~2.
By repeating the trial many times, we statistically evaluate the
probabilities $P(n_{1},n_{2};\phi_{j})$ of detecting $n_{1}$ and $n_{2}$
photons in modes~1 and~2, respectively, when the reference state
$|\phi_{j},s\rangle_{1}$ is injected.
From these probabilities, the coefficients $\{c_{k}\}$ can be estimated.

To simplify the problem, we set $s=1$ as a concrete example.
The unknown state can then be written as
$|\psi\rangle_{2} = c_{0}|\phi_{0},1\rangle_{2} + c_{1}|\phi_{1},1\rangle_{2}$,
and the goal of the experiment is to estimate $c_{0}$ and $c_{1}$.
Using the beam-splitter transformation, the output state is expressed as
\begin{equation}
\hat{B}
|\phi_{j},1\rangle_{1}
|\psi\rangle_{2}
=
\frac{1}{2}
\sum_{n=0}^{1}
\sum_{m=0}^{1}
e^{in\phi_{j}}
(c_{0}e^{im\phi_{0}} + c_{1}e^{im\phi_{1}})
\frac{1}{\sqrt{n!m!2^{\,n+m}}}
(\hat{a}_{1}^{\dagger}-\hat{a}_{2}^{\dagger})^{n}
(\hat{a}_{1}^{\dagger}+\hat{a}_{2}^{\dagger})^{m}
|0\rangle_{1}|0\rangle_{2}.
\end{equation}
The detection probabilities $P(n_{1},n_{2};\phi_{j})$ for observing
$|n_{1}\rangle_{1}|n_{2}\rangle_{2}$ are then given by
\begin{eqnarray}
P(0,0;\phi_{j})
&=&
\frac{1}{4}|c_{0}+c_{1}|^{2}, \nonumber \\
P(0,1;\phi_{j})
&=&
\frac{1}{8}
|c_{0}e^{i\phi_{0}}+c_{1}e^{i\phi_{1}}-(c_{0}+c_{1})e^{i\phi_{j}}|^{2}, \nonumber \\
P(1,0;\phi_{j})
&=&
\frac{1}{8}
|c_{0}e^{i\phi_{0}}+c_{1}e^{i\phi_{1}}+(c_{0}+c_{1})e^{i\phi_{j}}|^{2}, \nonumber \\
...
&&
.
\end{eqnarray}

We now consider the problem more concretely. Neglecting the global phase and
imposing the normalization condition, the coefficients of $|\psi\rangle_{2}$ can be written as
\begin{equation}
c_{0}=r,\qquad
c_{1}=\sqrt{1-r^{2}}\,e^{i\theta},\qquad
0\le r\le 1,\qquad
0\le\theta<2\pi.
\end{equation}
Because $s=1$, we have $\phi_{1}=\pi$ with $\phi_{0}=0$. Moreover, we set
$\phi_{j}=0$. Then we obtain
\begin{equation}
P(0,0;0)
=
\frac{1}{4}\left|r+\sqrt{1-r^{2}}\,e^{i\theta}\right|^{2},
\end{equation}
\begin{equation}
P(0,1;0)
=
\frac{1}{2}(1-r)^{2}.
\end{equation}
Thus, $r$ and $\theta$ can be determined from $P(0,0;0)$ and $P(0,1;0)$.

For arbitrary $s$, the number of probabilities
$P(n_{1},n_{2};\phi_{j})$ for $0\le n_{1},n_{2}\le s$ with $(n_{1},n_{2})\neq (0,0)$ and
$\{\phi_{0},\phi_{1},\ldots,\phi_{s}\}$ is $[(s+1)^{2}-1](s+1)$.
By contrast, the number of unknown real parameters in the coefficients
$\{c_{k}\}$ of $|\psi\rangle_{2}$ is
$2(s+1)-2 = 2s$ after neglecting the global phase and imposing the normalization condition
$\langle\psi|\psi\rangle=1$.
Since $[(s+1)^{2}-1](s+1)> 2s$ for $s\ge 1$, the values of $\{c_{k}\}$ can always be
estimated from the set $\{P(n_{1},n_{2};\phi_{j})\}$.

\section{\label{section-discussion}Discussion}

The procedure examined in Sec.~\ref{section-application-phase-measurement} can be regarded as a phase-estimation protocol for
the given unknown state.
The phase-estimation method discussed in
Sec.~\ref{section-application-phase-measurement} is not efficient from a
practical viewpoint. Addressing this issue is an important direction for
future work.

Because each Pegg-Barnett eigenstate is an equally weighted superposition of
photon-number states, the photon-number variance is large even though the phase
is sharply defined. These features are consistent with the
uncertainty relation of number and phase.
Our analysis of the experimental setup shown in Fig.~\ref{figure03} for generating the Pegg-Barnett
eigenstates indicates that it is difficult to realize such eigenstates in a
Hilbert space of large dimension, i.e., $s \gg 1$,
because the generation probability decreases rapidly with increasing $s$.
This observation is compatible with
the well-known argument by Susskind and Glogower that no Hermitian phase
operator exists for photons.
In short, the difficulty in generating the Pegg-Barnett phase-operator eigenstates arises from the fundamental obstacle of preparing superpositions of photon-number states with equally-weighted amplitudes in a high-dimensional Hilbert space.

\section{\label{section-conclusion}Conclusions}

In this paper, we began by computing the Wigner functions and negativity volumes of
the Pegg-Barnett phase-operator eigenstates and showed that these
eigenstates exhibit non-Gaussianity.
According to results of our numerical computation, the negativity volume diverges as the dimension of the Hilbert space on which the eigenstate defined increases.
We then presented a quantum-optical circuit that generates the eigenstates.
From this implementation, we identified the single-photon detection as the origin of the non-Gaussianity of the eigenstates.
Next, by introducing imperfect single-photon detectors with non-unit efficiency
into the circuit, we examined how the detector efficiency affects the generation probabilities, the fidelities, and the negativity volumes of the output states that approximate the Pegg-Barnett phase-operator eigenstates.
We also considered an experimental method for determining the
coefficients of an unknown superposition of the Pegg-Barnett phase-operator
eigenstates. 

\section*{Acknowledgment}

This paper is based on results obtained from a project,
JPNP25014,
commissioned by the New Energy and Industrial Technology
Development Organization (NEDO).

\appendix
\renewcommand{\thesection}{Appendix \Alph{section}}

\section{\label{Appendix-A}Procedure for expressing the
\\
displacement‑operator parameters as
\\
functions of the squeezing parameter}

Applying the beam splitters and the displacement operators to the two-mode squeezed vacuum state as shown in Fig.~\ref{figure03}, we obtain the state,
\begin{eqnarray}
|\Psi\rangle
&=&
\hat{D}_{1}(\alpha_{1})
\hat{D}_{2}(\alpha_{2})
\cdots
\hat{D}_{s}(\alpha_{s})
\hat{B}_{s-1,s}
\hat{B}_{s-2,s}
\cdots
\hat{B}_{1,s} \nonumber \\
&&
\times
\sqrt{1-q^{2}}
\sum_{n=0}^{\infty} q^{n}
|0\rangle_{1}
|0\rangle_{2}
\cdots
|0\rangle_{s-1}
|n\rangle_{s}
|n\rangle_{\mathrm{A}}.
\label{state-circuit-output-0}
\end{eqnarray}
A single-photon detection in each of modes $1, 2, ..., s$ projects mode A onto the state
\begin{equation}
\sum_{j=0}^{s} c_{j} q^{j} |j\rangle_{\mathrm{A}} + O(q^{s+1}),
\label{projected-state-mode-A-0}
\end{equation}
where
\begin{eqnarray}
c_{0} &=& \alpha_{1}\alpha_{2}...\alpha_{s}, \nonumber \\
c_{1} &=& \frac{1}{\sqrt{s}}(
\alpha_{1}\alpha_{2}...\alpha_{s-1}
+
\alpha_{1}\alpha_{2}...\alpha_{s-2}\alpha_{s}
+
...+\alpha_{2}\alpha_{3}...\alpha_{s-1}\alpha_{s}
), \nonumber \\
c_{2} &=& \frac{\sqrt{2}}{s}(
\alpha_{1}\alpha_{2}...\alpha_{s-3}\alpha_{s-2}
+
\alpha_{1}\alpha_{2}...\alpha_{s-3}\alpha_{s-1}
+
\alpha_{1}\alpha_{2}...\alpha_{s-3}\alpha_{s}
+
...
+
\alpha_{3}\alpha_{4}...\alpha_{s}
), \nonumber \\
...&&\nonumber \\
c_{s-1} &=& f(s)
(\alpha_{1}+\alpha_{2}+...+\alpha_{s}), \nonumber \\
c_{s} &=& f(s),
\label{coefficients-projected-state-mode-A-0}
\end{eqnarray}
\begin{equation}
f(s)
=
\sqrt{s!}\prod_{j=2}^{s}
j^{-j/2}(j-1)^{(j-1)/2}
\quad
\mbox{for $s=3,4,...$.}
\end{equation}
As shown in Eq.~(\ref{coefficients-projected-state-mode-A-0}),
$c_{1}$ is proportional to the sum of all symmetric permutations of $(s-1)$ elements taken from $\{\alpha_{j}\}$,
and $c_{2}$ is proportional to the sum of all symmetric permutations of $(s-2)$ elements drawn from the set $\{\alpha_{j}\}$.

By imposing the condition
\begin{equation}
c_{0} = q c_{1} = q^{2} c_{2} = ... = q^{s-1} c_{s-1} = q^{s} c_{s}
= \mathrm{Constant},
\label{relations-coefficients-projected-state-mode-A-0}
\end{equation}
the parameters $\{\alpha_{j}\}$ are obtained as solutions of an $s$-th degree polynomial equation
\begin{eqnarray}
&&
(x-\alpha_{1})(x-\alpha_{2})...(x-\alpha_{s}) \nonumber \\
&=&
x^{s}
- q x^{s-1}
+...
+(-q)^{s-2}\frac{s f(s)}{\sqrt{2}} x^{2}
+(-q)^{s-1}\sqrt{s}f(s)x
+(-q)^{s}f(s) \nonumber \\
&=&
0.
\label{s-th-degree-polynomial-equation}
\end{eqnarray}
If $s\leq 4$, closed-form expressions of solutions of Eq.~(\ref{s-th-degree-polynomial-equation}) are available,
and
we can obtain $\{\alpha_{j}\}$ analytically as functions of $q$.
However, for $s\geq 5$, no closed-form expressions of solutions exist for Eq.~(\ref{s-th-degree-polynomial-equation}),
so that we must compute the $s$ solutions numerically. 
By solving Eq.~(\ref{s-th-degree-polynomial-equation}), we obtain $\{\alpha_{j}\}$ as functions of $q$.

\end{document}